\documentclass [dvips,fleqn] {www2007-submission}

\usepackage {amsmath,float,alltt,color}

\usepackage {graphicx}

\floatstyle {ruled}
\newfloat {algorithm}{tbhp}{lop}
\floatname {algorithm}{Algorithm}

\newcommand\seclabel [1] {\label {sec: #1}}
\newcommand\secref [1] {Section~\ref {sec: #1}}
\newcommand\sseclabel [1] {\label {ssec: #1}}
\newcommand\ssecref [1] {Subsection~\ref {ssec: #1}}
\newcommand\figlabel [1] {\label {fig: #1}}
\newcommand\figref [1] {Figure~\ref {fig: #1}}
\newcommand\tbllabel [1] {\label {tbl: #1}}
\newcommand\tblref [1] {Table~\ref {tbl: #1}}

\newcommand\rsv [1] {\ensuremath {\mathbf {#1}}}

\newcommand\clustering [0] {\ensuremath {{\cal C}}}
\newcommand\set [1] {\ensuremath {\left\{#1\right\}}}

\newcommand\DQ [3] {\ensuremath {\Delta Q^{#1}_{#2,#3}}}
\newcommand\mixiGraph [1] {\ensuremath {{G_{\mathrm {mixi}}^{#1}}}}

\newcommand\Ratio[2]{\ensuremath{\mathit {ratio}(#1, #2)}}
\newcommand\E[2]{}
\def\E(#1,#2){\ensuremath {#1 \cdot 10^{#2}}}

\begin {document}

\title {Finding Community Structure in Mega-scale Social Networks}
\numberofauthors {2}
\author {
  \alignauthor Ken Wakita\\
      \affaddr {Tokyo Institute of Technology} \\
      \affaddr {2-12-1 Ookayama, Meguro-ku} \\
      \affaddr {Tokyo 152-8552, Japan} \\
      \email {wakita@is.titech.ac.jp}
  \alignauthor Toshiyuki Tsurumi\\
      \affaddr {Tokyo Institute of Technology} \\
      \affaddr {2-12-1 Ookayama, Meguro-ku} \\
      \affaddr {Tokyo 152-8552, Japan} \\
      \email {tsurumi2@is.titech.ac.jp}}

\maketitle

\begin {abstract}
Community analysis algorithm proposed by Clauset, Newman, and Moore (CNM algorithm) finds community structure in social networks.  Unfortunately, CNM algorithm does not scale well and its use is practically limited to networks whose sizes are up to 500,000 nodes.  The paper identifies that this inefficiency is caused from merging communities in unbalanced manner.  The paper introduces three kinds of metrics (\emph {consolidation ratio}) to control the process of community analysis trying to balance the sizes of the communities being merged.  Three flavors of CNM algorithms are built incorporating those metrics.  The proposed techniques are tested using data sets obtained from existing social networking service that hosts 5.5 million users.  All the methods exhibit dramatic improvement of execution efficiency in comparison with the original CNM algorithm and shows high scalability.  The fastest method processes a network with 1 million nodes in 5 minutes and a network with 4 million nodes in 35 minutes, respectively.  Another one processes a network with 500,000 nodes in 50 minutes (7 times faster than the original algorithm), finds community structures that has improved modularity, and scales to a network with 5.5 million.
\end {abstract}




\section {Introduction}

Research of complex networks attracts interests of broad scientific disciplines.  Examples of complex networks include World Wide Web (WWW), citation networks, human activities on the Internet (e.g., exchange of emails, social networking system, consumption behavior on the e-commerce, and Web-log track-back network), physical phenomena, and biochemical networks among many others.

Finding community structure in networks is an important first step to grasp inherent complex structure of social networks.  Due to ever expanding use of digital networks, traces of global human activities have become available in digital forms.  There are many research activities that attempt to define the notion of communities and propose community analysis algorithms \cite {k98:HITS,gkr98:HITS_analysis,krrt99:trawling,dh99:companion,tk01:companion--,m+01:HITS+,wh03:linear_time,ng04:modularity,cnm04:very_large_network,r+04:defining_identifying,chwm04:block-level,ow05:graph_partition}.

We implemented a fast community analysis algorithm proposed by Clauset, Newman, and Moore \cite {cnm04:very_large_network} (CNM algorithm) and applied it to analyze various subsets of an acquaintance relationship network obtained from a social networking system (SNS).  The algorithm performs well for a mid-scale subset of the network that consists of less than 500,000 users.  However, the algorithm was incapable to analyze larger networks.

We observed that merging communities of unbalanced sizes has great impact on computational efficiency of CNM algorithm.  From this observation it was expected that merging communities in a balanced manner will improve the efficiency of the algorithm.  In this paper, we introduce the notion of \emph {consolidation ratio}, which is a measure of balancedness of the community pairs, and use it as well as \emph {modularity} as means to find next pair of communities to merge into a larger one.

The paper presents three types of consolidation ratio.  Three flavors of CNM algorithms, each of which incorporates one of those consolidation ratio, were built.  They are implemented as a single-threaded Java program and were tested using as data sets various subsets of a SNS network that hosts 5.5 million users.  The fastest program finds community structure in a network of 1 million nodes in 5 minutes.  Computational efficiency and scalability of the proposed algorithm, and quality of the generated community structures are discussed in detail.

The structure of the paper is as follows: \secref {related work} compares our work with other related research activities, \secref {clauset} explains the CNM algorithm and identifies the source of its performance inefficiency, \secref {algorithm} introduces a heuristics that makes use of consolidation ratio, \secref {evaluation} evaluates the proposal, and \secref {summary} concludes the paper.


\section {Related Work}
\seclabel {related work}

Analysis of community structures of social and cyber networks is an effort to find cyber-communities.  We believe that such found cyber-communities support reasoning about structure, nature, and dynamics of real-communities.  Many community analysis techniques have been proposed by researchers of broad discipline.  There are two types of algorithms that are designed for this purpose.  One type takes a graph and one or more seed node(s), and gives a community structure that includes the seed node(s) \cite {k98:HITS,dh99:companion,tk01:companion--,m+01:HITS+}.  This type of community analysis algorithm is widely used for analysis of WWW link structure.  In WWW link analysis, Web pages or Web sites are modeled as nodes and hyper-links are treated as edges, forming a huge directed graph.

`HITS' algorithm \cite {k98:HITS,gkr98:HITS_analysis} proposed by Kleinberg focuses on two types of characteristic structures called authorities and hubs that are defined in mutually recursive manner.  A Web page given a higher authority value is regarded as an authoritative page.  It is referenced from many hub pages which in turn collect many links to authoritative pages.  HITS algorithm assigns an authority value and a hub value to each Web page in an iterative process.  Link structures formed by authorities and hubs can be understood as cores of inter-related community structures.

Dean and Henzinger used HITS algorithm to build a new Web search engine called `Companion' \cite {dh99:companion}.  Unlike standard keyword-based search engines, Companion takes Web pages of interest for the user and performs a Web link analysis to find a set of Web pages whose contents are closely related with each other.  Toyoda and Kitsuregawa improved the performance of Companion's link analysis and proposed an improved version called `Companion--'.  Companion-- visually addresses internal structure of the Web community \cite {tk01:companion--}.

Another type of community analysis algorithms takes a graph and divide it into a set of densely connected subgraphs \cite {k98:HITS,krrt99:trawling,flg00:efficientIdentification,f+02:self-organization-identification,wh03:linear_time,ng04:modularity,cnm04:very_large_network,chwm04:block-level,ow05:graph_partition}.  Various notions of communities have been proposed.  Some work ``defines'' communities by the algorithm.  Kumar and others formulated graph partition problem as finding minimum complete bipartite subgraphs.  Flake and others gave a concise definition of cyber-communities based on graph-theoretic foundation \cite {flg00:efficientIdentification,f+02:self-organization-identification} and proved that community analysis falls into maximum-flow, minimum-cut problem.  Newman and Girvan proposed a measure called \emph {modularity}, which is a quantitative measure of quality of graph partitioning \cite {ng04:modularity}.  A fast algorithm that finds a community structure in a bottom-up manner, greedily maximizing on modularity was presented in \cite {cnm04:very_large_network}.  Our research is based on this work.


\section {CNM Algorithm}
\seclabel {clauset}

Newman and Girvan attempt to measure the quality of network clustering by means of \emph {modularity} \cite {ng04:modularity}.  Their algorithm (CNM algorithm) is a bottom-up greedy optimization that continuously finds and merges pair of communities trying to maximize modularity of the community structure \cite {cnm04:very_large_network}.  This section briefly presents the notion of modularity, an outline of CNM algorithm, and addresses its computational inefficiency.

\subsection {Modularity}

\emph {Modularity} of network's community structure is a quantitative measure of the quality of \emph {clusterings} (i.e., a graph partitioned into a set of subgraphs) \cite {ng04:modularity}.  It can be used to compare the quality of different clusterings of the same network.  It is desirable that members of a community have a dense intra-community links and small number of links connected to members of other communities.  This idea is embedded in the formulation of modularity as explained subsequently.

Let $G = (V, E)$ be a undirected graph that represents a social network.  For example, an acquaintance network of a SNS can be represented by $(U, F)$, where $U$ is a set of users and $F$ represents friendship (if users $u_1$ and $u_2$ are friends then $(u_1, u_2) \in F)$).  Adjacency matrix $A$ is another way to represent edges:
\[
  A_{vw} =
    \begin {cases}
      1 & (v, w) \in E \\
      0 & \text {otherwise}.
    \end {cases}
\]
It can be used to define the number of total edges ($m = \sum_{v, w \in V} A_{vw} / 2$) and the degree of a node $v$ ($k_v = \sum_{w \in V} A_{vw}$).

A clustering (\clustering) of $G$ into a set of communities is a partitioning of nodes $V$ into its subsets:
\[ \clustering = \set {c_1, c_2, \ldots},\,\,
   c_i \cap c_j = \emptyset \,\, (i \neq j),\,\,
   \bigcup_{c_i \in \clustering} c_i = V
\]

Proportion of edges that link members of communities $c_i$ and $c_j$ in the whole graph is given by $e_{ij}$.  Likewise proportion of $c_i$'s edges in the whole graph is given by $a_i$:
\begin {align*}
e_{ij} & = \sum_{v \in c_i, w \in c_j}A_{vw} / 2m\\
a_i & = \sum_{v \in c_i} k_v / 2m.
\end {align*}

Definition of modularity as given below states that communities in a good clustering of a graph $G$ has dense intra-community links and less inter-community links:
\[ Q(G, \clustering) = \sum_i(e_{ii}-a_i^2). \]

\subsection {Algorithm}

Newman and Girvan presented a greedy community analysis algorithm that optimize on modularity.  Later, Clauset, Newman, and Moore proposed a more efficient algorithm (\emph {CNM algorithm}) that works the same as the former proposal in principle but incorporates sophisticated data structures \cite {cnm04:very_large_network}.

The algorithm starts from a totally unclustered situation, where each node in a graph forms a singleton community.  Then computed is for each pair of communities, expected improvement of modularity when they merge:
\[ \DQ \clustering {c_i} {c_j} =
   Q(G, \clustering - c_i - c_j + (c_i \cup c_j)) -
   Q(G, \clustering). \]

The algorithm repeatedly chooses a community pair that gives the maximum $\Delta Q$ value and merges them into a new community (Algorithm~\ref {algorithm: Clauset+}).  During the merge process, $\Delta Q$ values of the communities that adjoin the new community needs to be updated.  Because the number of community pairs in the clustering decreases monotonously, the algorithm eventually stops when there remains no community pairs to merge.

\begin {algorithm}
\begin {quote}
$\clustering := \{ v \in V | \{ v \} \};$ \medskip \\
\rsv {function} join($c_i, c_j$) \{\\
\quad \rsv {return} $\clustering - c_i - c_j + (c_i \cup c_j)$; \\
\} \medskip\\
\rsv {procedure} updateDeltaQ() \{\\
  \quad $\forall c_i, c_j \in \clustering.\\
    \qquad \DQ \clustering i j := Q(G, \text {join}(c_i, c_j)) - Q(G, \clustering);$\\
\} \medskip\\
\rsv {while} (\rsv {true}) \{\\
  \quad updateDeltaQ(); \\
  \quad Find $(c_i, c_j) \in \clustering^2 $ that has maximum \DQ \clustering {c_i} {c_j}.\\
  \quad \rsv {if} ($\max(\DQ \clustering {c_i} {c_j} < 0$) \rsv {break};\\
  \quad \clustering := join$(c_i, c_j)$;\\
\}
\end {quote}
\caption {An outline of the algorithm proposed by Clauset et al \protect\cite {cnm04:very_large_network}}
\label {algorithm: Clauset+}
\end {algorithm}

CNM algorithm uses two data-structures to find a community pair with maximum $\Delta Q$ value: (1) a balanced binary tree (or heap tree) of community pairs $(c_i, c_j)$ and (2) a max heap (or priority heap) of community pairs that is sorted by $\DQ \clustering {c_i} {c_j}$.  They achieve logarithmic order of computational cost for removal and insertion of a community pair, and finding a community pair $(c_i, c_j)$ with maximum $\Delta Q$ value for a given $c_i$.  For each community, the community pair with maximum $\Delta Q$ value are stored in a system-wide max heap.

By using these data structures, search for the community pair with the largest $\Delta Q$ value is performed in two stages.  Firstly, each community searches in its max heap for the pair with the largest $\Delta Q$ among its community pairs and stores it in a system-wide max heap that is used in the second stage.  Elements in the system-wide max heap are candidates of the community pair who has system-wide maximum $\Delta Q$ value.  When all the candidates are stored in the system-wide max heap, the pair with system-wide maximum $\Delta Q$ value can be easily found.\label {two staged}

Newman and Girvan showed that update of $\DQ \clustering {c_i} {c_j}$ for a community pair $(c_i, c_j)$ needs to be performed only when either $c_i$ or $c_j$ merges.  Also update of $\DQ \clustering {c_i} {c_j}$ is a simple arithmetics using its neighbors' past $\Delta Q$ values. Clauset and others have applied this algorithm to several real world social networks including purchase transactions offered by Amazon which contains more than 400,000 nodes and 2 million edges.\footnote {http://www.amazon.com/}

\subsection {Performance inefficiency}
\sseclabel {performance inefficiency}

The authors have programed CNM algorithm and attempted to analyze an acquaintance network of an SNS called ``mixi\footnote {mixi (http://mixi.jp/) is the largest invitation-based SNS in Japan.}'' that hosted about one million users in October 2005.  The experiment was performed on a PC (Intel Xeon 2.80GHz, L2 cache = 2MB, Memory = 4GB).  However, in spite of the good scalability as advertised in \cite {cnm04:very_large_network}, the authors have found it was impractical to analyze this mega-scale social network using CNM algorithm.  The experiment was stopped after a week when less than 10\% of the whole analysis was finished.  Yuta and others has conducted similar experiment on earlier mixi network on Linux running on Pentium IV 2.8 GHz with 1GB memory and states that community analysis of an SNS network of 360,000 users using CNM algorithm took six hours \cite {yof06:mixiAnalysis,ynf07-GapInCommunitySize}.

\begin{figure}[htbp]
  \centerline {\includegraphics*[width=0.95\linewidth]{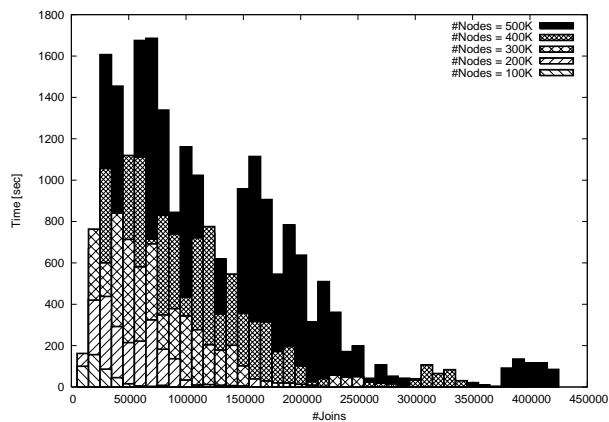}}
  \caption {Analysis time required for networks with various scales (100K, 200K, \ldots, 500K nodes).  Each bar represents time required for merging 10,000 community pairs.}
  \figlabel {Clauset+ computation broken down per 10,000 joins.}
\end{figure}

To figure out the performance bottleneck of CNM algorithm, we conducted community analysis on a various subsets of mixi SNS network.  The mixi SNS gives each user an ID number starting from ``1'', in the order of user registration.  Therefore, the mixi SNS network can be represented by a graph $\mixiGraph{} = (U, F)$, where $U = \{ 1, 2, \ldots \}$ is the set of user IDs and $F \subset U \times U$ is a set of acquaintance relationship, namely $(i, j) \in F$ if and only if two users identified by $i$ and $j$ are friends.  We built a subset of mixi acquaintance graph $\mixiGraph n$ as follows:
\begin {align*}
  \mixiGraph n = (U(n), F \cap (U(n) \times U(n))) \\
  \text {where } U(n) = \{ u \in U | u \le n \}
\end {align*}

\figref {Clauset+ computation broken down per 10,000 joins.} illustrates time required for community analysis of various subsets of the social network: $\mixiGraph {100K},$ $\mixiGraph {200K},$ $\mixiGraph {300K},$ $\mixiGraph {400K},$ and $\mixiGraph {500K}$.  Each bar of the graph depicts time required to perform 10,000 merges of community pairs.  For example, in case of $\mixiGraph {500K}$ (black bars), 427,794 merges are performed and the third 10,000 merges took about 1,600 seconds.

For each data set, most of the computation time is consumed for the first half of the merging process and computation time decreases dramatically for the latter half.  For example, in case of $\mixiGraph {500K}$, merging 10,000 communities takes less than 200 seconds after 250,000 communities are merged.

The gross area of each pattern is the elapsed time of respective subset of the network (Elapsed time for \mixiGraph n is compared with our proposal in \figref {comparison/etime-size} on page~\pageref {fig: comparison/etime-size}).  In this experiment, we can approximate the elapsed time for analysis of \mixiGraph {n} by $T(n) \approx 1.5 \cdot 10{-8} x^{2.13 \pm 0.104}$.

\cite {cnm04:very_large_network} estimates the computational complexity of CNM algorithm to be $O(md \log n),$ where $n$ and $m$ are numbers of nodes and edges, respectively, and $d$ is the height of \emph {dendrogram}\footnote {A dendrogram is a binary tree that represents a history of merge process.  If a pair of nodes $(c_i, c_j)$ are merged into a new community $c_k$, the dendrogram for $c_k$ will be a binary tree whose subtrees are dendrograms for $c_i$ and $c_j$.}.  It also discusses in a sparse network $m$ and $d$ can be approximate by $n$ and $\log n$, respectively and that computational complexity will be $O(n\log^2n)$ for social networks.  This discussion and the above mentioned super quadratic computational cost observed in our experiment contradict.  Investigation of the structure of the dendrogram suggests that $d \approx \log n$ does not hold for the analysis of mixi SNS network.

\begin{figure}[htbp]
  \centerline {\includegraphics [width=0.95\linewidth] {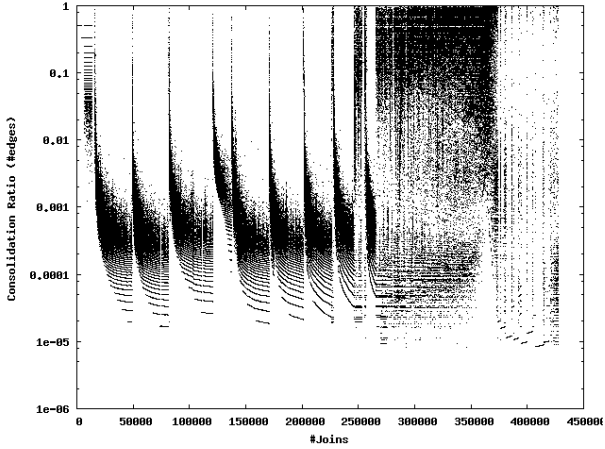}}
  \caption {Consolidation ratio of each merge step illustrated in a partially log-scale chart.}
  \figlabel {clauset/ratio-join}
\end{figure}

Then the authors carefully observed a merge logs that record how community pairs are merged into larger ones.  The merge logs suggested that among huge number of communities only a small portions are growing fast, merging in many tiny communities.  Because of this phenomenon, a huge unbalanced dendrogram was constructed.

This phenomenon can be clearly seen in \figref {clauset/ratio-join} which presents unbalancedness of merge steps are through out the progress of community analysis for \mixiGraph {500K}.  For this purpose, we have defined the notion of \emph {consolidation ratio} of community merge, which is defined as follows:
\[
  \Ratio {c_i}{c_j} = \min(|c_i|/|c_j|, |c_j|/|c_i|).
\]

\figref {clauset/ratio-join} plots, for $n$-th merge step, $c_k := \text {join}(c_i, c_j)$, $(n, \Ratio {c_i} {c_j})$, where the size of a community ($|c|$) is measured in terms of the number of its links to other communities.  In this figure, we can see growth of some eight large communities in the first half of the community analysis.  We can conclude that unbalanced growth of large communities is the primary cause of performance degradation when CNM algorithm is applied to our dataset.

Unbalanced merging process, makes the height of the dendrogram grow more or less proportionally to its size and leads to degrade the computational efficiency to $O(n^2 \log n)$.


\section {Algorithm}
\seclabel {algorithm}

In the previous section, we have seen the cause of the inefficiency of CNM algorithm.  In this section, we present a data structure and three types of heuristics that dramatically improve computational efficiency of CNM algorithm.

\subsection {Data structure}

\begin {figure}
  \centerline {\includegraphics [width=0.95\linewidth]
              {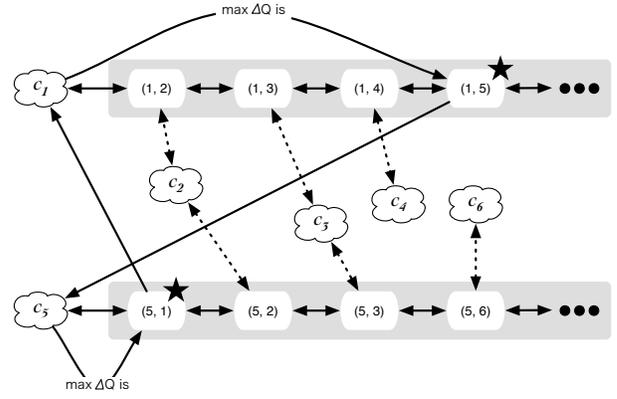}}
  \caption {Our implementation of communities.  A community maintains a link to its neighboring communities in a list of community pairs and a pair that has maximum $\Delta Q$ value.}
  \figlabel {data structure: before}
\end {figure}

In CNM algorithm, heavy operations are performed when it finds for the community pair that has the maximum $\Delta Q$ value and when merging communities.  We have replaced balanced binary trees and max heaps, originally suggested in \cite {cnm04:very_large_network} by a doubly-linked list that is sorted in the order of community ID.

Each community $c_i$ in our system has a data structure to store references to neighboring communities which is represented by a list of pairs of communities (see \figref {data structure: before}).  The list is sorted by the order of \emph {Community ID}.  For example, a community $c_1$ that links to communities $c_2$, $c_3$, $c_4$, $c_5$, \ldots is represented by a community object that has a list of community pairs $\set { (1, 2), (1,3), (1,4), (1,5), \ldots }$.  A community pair has references to the communities it belongs to.  For example, in \figref {data structure: before}, community pair $(c_1, c_2)$ has links pointing at communities  $c_1$ and $c_2$.  Merging two communities effectively is a process of merging their community pairs, eliminating duplicates and updating their $\Delta Q$ values.  By the use of sorted lists, merging can be accomplished in linear order to the number of community pairs.

Similarly to \cite {cnm04:very_large_network}, each community nominates its \emph {largest community pair} (the pair in its community pair list that has the largest $\Delta Q$ value) to be stored in the system-wide max heap.  This technique allows for efficient retrieval of \emph {maximum community pair} (the pair of communities that has the largest $\Delta Q$ value, system-wide).  For this purpose, each community maintains a link to the largest pair of communities among members of its list.  \figref {data structure: before} marks the largest community pair of communities by black stars ($\star$'s) and links to the largest community pairs by ``\emph {max $\Delta Q$ is}'' links.  When two communities merge, the ``\emph {max $\Delta Q$ is}'' link for the new community can simply be found because anyway we need to scan all the community pairs to merge them (\figref {data structure: after}).

\begin {figure}
  \centerline {\includegraphics [width=0.95\linewidth]
              {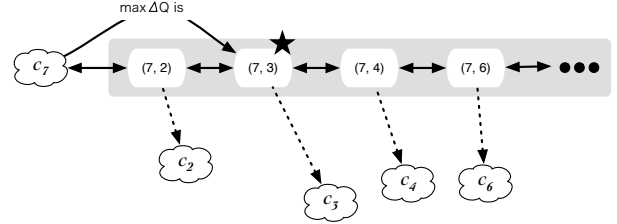}}
  \caption {Merge of $c_1$ and $c_5$ in \figref {data structure: before} produced a new community $c_7$.  During the merge, community pairs for the merged updating their $\Delta Q$ values.}
  \figlabel {data structure: after}
\end {figure}

The use of ``\emph {max $\Delta Q$ is}'' link, however, introduces an unpleasant problem.  When communities $c_i$ and $c_j$ merge and $\Delta Q$ value of community pair $p = (c_i, c_k)$ is updated, we need to maintain the integrity of $c_k$ such that its ``\emph {max $\Delta Q$ is}'' link points to the truly largest community pair in $c_k$'s list.

\begin {itemize}

\item If $p$ is not the largest community pair of $c_k$ (or more casually $p$ is not marked by a black star) and its $\Delta Q$ value decreases, nothing is needed.

\item If $p$ is not the largest community pair of $c_k$ and its $\Delta Q$ value increases, we need to compare it with $c_k$'s $\Delta Q$.  If the updated value is larger, the ``\emph {max $\Delta Q$ is}'' link is arranged to point to $p$ (or more casually, we remove a black star from $c_k$'s former largest community pair and put it to $p$).

\item If $p$ is $c_k$'s largest community pair and its $\Delta Q$ value increases, nothing is needed.

\item (The Worst case) If $p$ is $c_k$'s largest community pair and its $\Delta Q$ value decreases, we do not have a convenient means to tell if it remains the largest or not.  In this case, we scan all the community pairs of $c_k$ and find the largest one.

\end {itemize}

The reader may fear a scenario, where the last case is taken most of the time.  However, we believe it is not the case.  The $\Delta Q$ quantity for the community pairs depends on the number of neighboring communities that those pair have.  If the search process for community structure follows the \emph {preferential attachment law}\cite {ba99:scaleFreeNetwork}, it is expected that there exists a heavily linked pair in each community's list and its $\Delta Q$ is superior to those of other pairs'.  In such situation it would be very difficult for others to compete with the largest community pair.  If this optimistic anticipation is guaranteed, the update of $\Delta Q$ is performed in a unit cost for each community pair.

In summary, arranging a set of community pairs in a list allows for fast merging cost ($O(m)$ time), fast retrieval of the community pair with maximum $\Delta Q$ value ($O(1)$ time), and hopefully fast updates of $\Delta Q$ values for the community pairs ($O(m)$ time), where $m$ stands for the number of community pairs.

\subsection {Heuristics based on consolidation ratio}

In \ssecref {performance inefficiency}, we have seen that the performance of the algorithm degraded from unbalanced growth of large communities.  If, in certain way, we could control the growth of communities so that they grow in a balanced manner, it is anticipated that the performance of the algorithm will improve remarkably.  To turn this idea into practice, we tested three flavors of CNM algorithm that incorporate heuristics based on three kinds of consolidation ratio.

\begin {algorithm}
\begin {quote}
  \rsv {function} $\Ratio {c_i}{c_j} \{\\
  \quad \rsv {return} \min(|c_i|/|c_j|, |c_j|/|c_i|)$; \\
  \}\medskip \\
  \rsv {while} (\rsv {true}) \{\\
  \quad updateDeltaQ(); \\
  \quad Find $(c_i, c_j) \in \clustering^2 $ \\
  \qquad that has maximum $\DQ \clustering {c_i} {c_j} \cdot \Ratio {c_i} {c_j}$.\\
  \quad \rsv {if} ($\max(\DQ \clustering {c_i} {c_j} < 0$) \rsv {break};\\
  \quad \clustering := join$(c_i, c_j)$;\\
  \}
\end {quote}
  \caption {Outline of the proposed algorithm.  The updateDeltaQ function remains the same as Algorithm~\ref {algorithm: Clauset+}.}
  \label {algorithm: Proposal}
\end {algorithm}

The structure of the algorithm remains the same as Algorithm~\ref {algorithm: Clauset+}.  The only difference resides in the valuation basis of community pairs.  Algorithm~\ref {algorithm: Clauset+} uses $\DQ \clustering {c_i} {c_j}$ while we use combination of both $\DQ \clustering {c_i} {c_j}$ and consolidation ratio ($\Ratio {c_i} {c_j}$).  This heuristics is designed so that it suppresses unbalanced merge of communities and leads to balanced growth of communities.

So far we have not defined how we measure the size of a community ($|c_i|$).  We have defined three different valuation of community size and developed three kinds of heuristics.

The first heuristics (\emph {HE}) measures the community size in terms of its degree (i.e., the number of edges linked to its neighboring communities or the length of its list of community pairs).  This heuristics was induced from the fact that the cost for merging communities is proportional to the number of their community pairs (see page~\pageref {two staged}).

The second heuristics (\emph {HE'}) was found accidentally when we were trying to implement HE.  As we have noted, the choice of the pair with largest $\Delta Q$ value is two staged.  For the first stage (selection of a candidate community pair), HE' ignores the size of a community and thus behaves equivalent to CNM algorithm.  On the other hand, for the second stage, where candidates pairs of maximum $\Delta Q$ is searched for, it measures community size in terms of its degree, like HE.  This weird heuristics, however, works faster than CNM algorithm and also it finds better clustering with respect to modularity.

The last heuristics (\emph {HN}) measures the size of community in terms of the number of its members.


\section {Evaluation}
\seclabel {evaluation}

This section presents results obtained from running four flavors of CNM algorithm, the original one proposed in \cite {cnm04:very_large_network} and three variations of Algorithm~\ref {algorithm: Proposal} that incorporate our heuristics (namely, HE, HE', and HN).

Four flavors of CNM algorithm, including the original one, are implemented using Java platform: Java 5.0, Java HotSpot Server VM (build 1.5.0\_06 b-05) with 3.2GB heap size.  The test was performed on a PC (CPU = Intel Xeon 2.80GHz, L2 Cache = 2MB, RAM = 4GB) running Linux (Red Hat Linux version 2.6.16).  Though Xeon comes with multiple cores, our Java program is single-threaded and makes use of no parallelism.

\subsection {Execution Efficiency}

\begin {figure}
  \centerline {\includegraphics [width=0.95\linewidth]
    {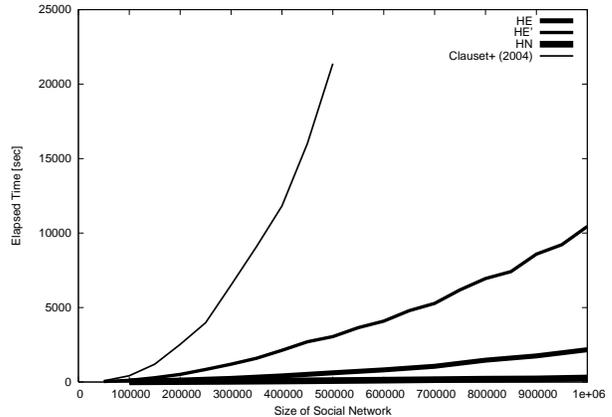}}
  \caption {Comparison of Elapsed Time}
  \figlabel {comparison/etime-size}
\end{figure}

\begin {table}
  \caption {Elapsed time (seconds)}
  \tbllabel {elapsed time}
  \centering
  \begin {tabular}{lrrrrr} \hline
        & 200K & 400K & 600K & 800K & 1M \\ \hline
    \textbf {Original}& 2,530 & 11,800 & NA & NA & NA \\
    \textbf {HE}  & 129  & 408  & 814  & 1470 & 2170 \\
    \textbf {HE'} & 511  & 2,130 & 4,090 & 7,410 & 10,400 \\
    \textbf {NE}  & 25.7 & 70.0 & 123  & 190  & 268 \\
    \hline
  \end {tabular}
\end {table}

Use of heuristics dramatically accelerates execution of community analysis.  We have applied four implementations to analysis of data sets $\mixiGraph {n}, (n \in \set {50K, 100K, \ldots, 1000K}$).  Results are presented in \figref {comparison/etime-size} and \tblref {elapsed time}.  The largest data set the original algorithm (Clauset+ (2004)) was possible to analyse is \mixiGraph {500K}.  It took about 5.9 hours.  The fastest heuristics was NE.  It processes \mixiGraph {1M} in less than five minutes.  Other heuristics, HE and HE', processes \mixiGraph {1M} in about 36 minutes and 3 hours, respectively.  They are slower than HE but still are practically usable, concerning the size of data sets.

\subsection {Consolidation Ratio}

\begin {figure}
  \begin {center}
	\includegraphics [width=0.95\linewidth]
	  {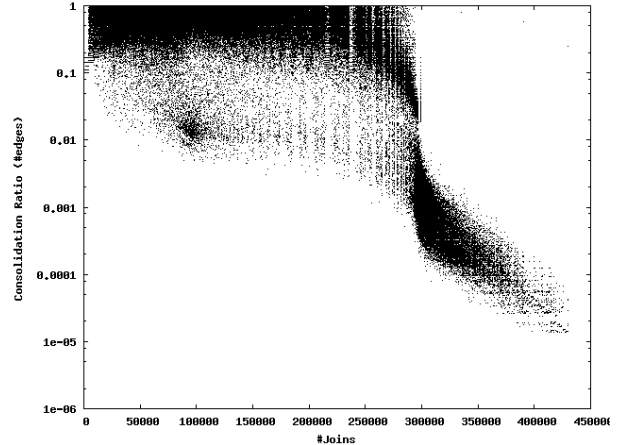} \\
    \textbf {(a) HE (\#edge ratio)} \bigskip \\
    \includegraphics [width=0.95\linewidth]
      {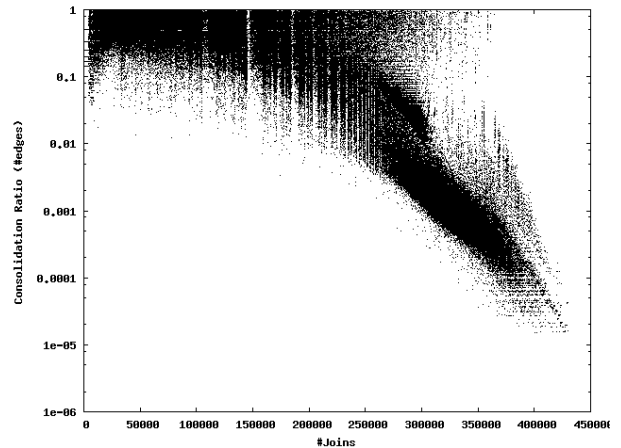} \\
    \textbf {(b) HE' (\#edge ratio with a bug)} \bigskip \\
    \includegraphics [width=0.95\linewidth]
      {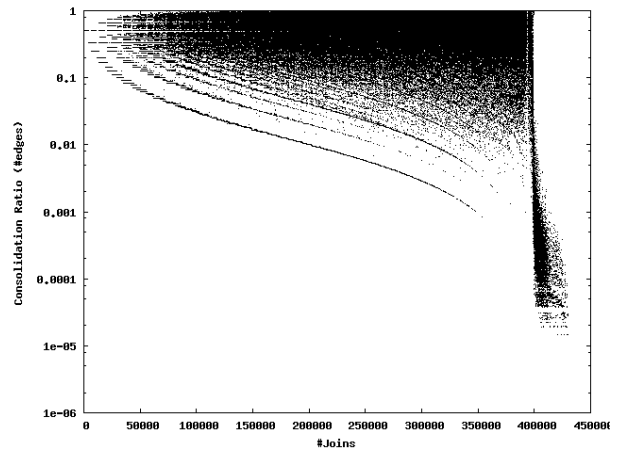} \\
    \textbf {(c) NE (\#node ratio)}
  \end {center}
  \caption {Consolidation ratio observed during analysis of \mixiGraph {500K}.}
  \figlabel {ratio-join-500K}
\end {figure}

Improvement of consolidation ratio of merged communities can explain the speed-up that we have seen previously.  \figref {ratio-join-500K}: (a)-(c) demonstrates consolidation ratios of merges of community pairs.  In \figref {clauset/ratio-join}, we have observed frequent unbalanced merges especially in the first half of community analysis.  Consolidation ratios were some 1:1,000 to 1:10,000.  In heuristics NE, the fastest one, for the most part of analysis consolidation ratios are kept better than 1:100 and most of the unbalanced merging are performed in the last stage of analysis.

We can observe similar phase-shift in heuristics HE but the phase-shift starts earlier than NE and phase transition is rather moderate.

In heuristics HE', it is difficult to observe a phase-shift that we have observed for NE and HE.  Consolidation ratios degrade slow as community analysis progresses.  As we will see shortly, this slow degradation of consolidation ratio seems to be a key issue in retaining higher modularity while achieving practical computational efficiency.

As we mentioned earlier, we can observe growth of several large communities in the earlier stage of the original algorithm (see \figref {clauset/ratio-join}).  In contrast, we can see many thin curves running from upper-left to central-right in \figref {ratio-join-500K}-(c).  It can be interpreted that multiple communities of different sizes are growing in a concurrent manner as community analysis progresses.  We believe concurrent growth of various communities gives more natural explanation to the community growth dynamics of a real SNS than than sequential development of large communities.

\begin {figure}
  \begin {center}
    \includegraphics [width=0.95\linewidth]
      {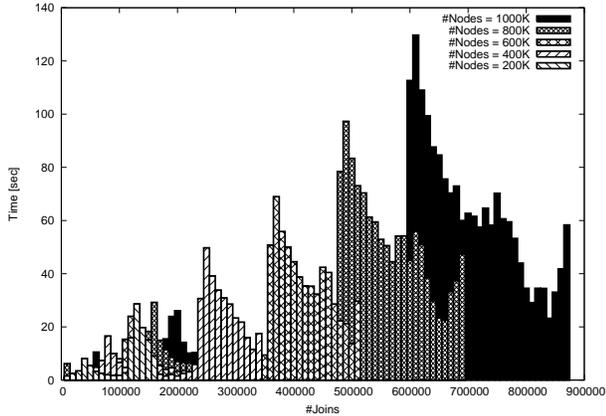} \\
    \textbf {(a) HE (\#edge ratio)} \bigskip \\
    \includegraphics [width=0.95\linewidth]
      {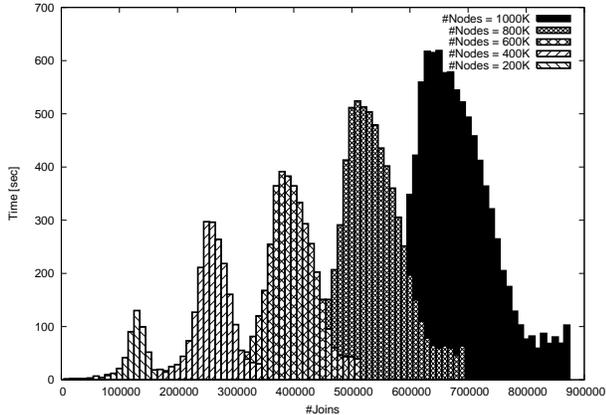} \\
    \textbf {(b) HE' (\#edge ratio with a bug)} \bigskip \\
    \includegraphics [width=0.95\linewidth]
      {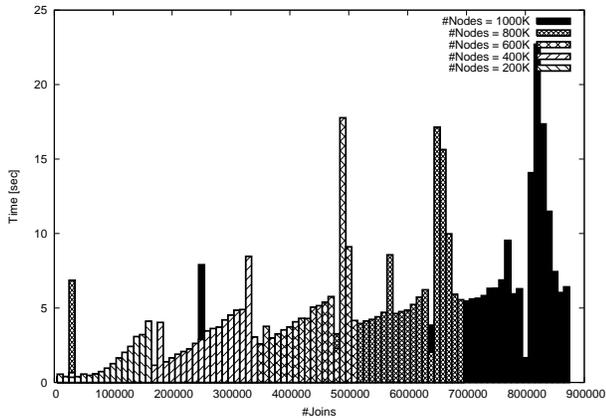} \\
    \textbf {(c) NE (\#node ratio)}
  \end {center}
  \caption {Analysis time required for networks of various scales.  Each bar represents time required for merging 10,000 community pairs.}
  \figlabel {time-join}
\end {figure}

The impact of the heuristics on improvement of analysis time can clearly be seen in \figref {time-join}: (a)-(c).  These charts presents time required for merging 10,000 community pairs.  The patterns painted on bars illustrate data sets of different scales ($\mixiGraph {n}, n \in \set {200K, 400K, 600K, 800K, 1000K}$).

Unlike \figref {Clauset+ computation broken down per 10,000 joins.}, computation cost is kept much cheaper up to the point when computational cost steeply increases.  The black bars stand for an experiment performed using \mixiGraph {1M}.  In this experiment, heuristics NE merges 10,000 communities in less than 7 seconds for the first 760K merges of communities  among 870K total merges.  It processes the heaviest part of the computation in less than 25 seconds, which is much smaller than heaviest computation cost performed in other heuristics, not to mention the original algorithm.

HE heuristics merges 10,000 communities in less than 5 seconds for the first 560K merges among 870K total merges.  In the computationally heavy part, it takes 60-130 seconds per 10,000 merges.

Merge cost of HE' heuristics is much higher than those of HE and NE.  In the computationally heavy part, it takes 100-650 seconds per 10,000 merges.

\subsection {Modularity}

\begin {figure}
  \centerline {\includegraphics [width=0.95\linewidth]
    {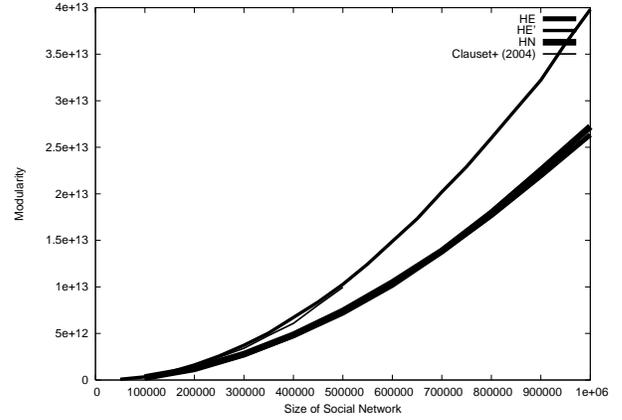}}
  \caption {Modularity of community structures resulted from community analysis performed on various scales.}
  \figlabel {comparison/modularity-size}
\end{figure}

\begin {table}
  \caption {Modularity}
  \tbllabel {modularity}
  \centering
  \begin {tabular}{lrrrrr} \hline
                  & 100K & 400K & 700K & 1M \\ \hline
    \textbf {Original} & \E(3.13, 11) & \E(6.08, 12) & NA & NA \\
    \textbf {HE}  & \E(2.61, 11) & \E(4.74, 12)
                  & \E(1.38, 13) & \E(2.63, 13) \\
    \textbf {HE'} & \E(3.38, 11) & \E(6.72, 12)
                  & \E(2.02, 13) & \E(3.98, 13) \\
    \textbf {NE}  & \E(2.66, 11) & \E(4.85, 12)
                  & \E(1.39, 13) & \E(2.72, 13) \\
    \hline
  \end {tabular}
\end {table}

It is our concern that use of heuristics reduces modularity of the resulting community structure.  \figref {comparison/modularity-size} and \tblref {modularity} presents modularity of the community structures obtained from the experiments.  The vertical shaft is $Q \cdot m^2$, where $Q$ is modularity as defined in \cite {cnm04:very_large_network} and $m$ is the number of edges in the graph.  In our implementation (including implementation of the original algorithm), we use $\Delta Q \cdot m^2$, instead of $\Delta Q$ because the former takes integer values and allows us to replace costly floating-point arithmetics by cheaper integer arithmetics.

To our surprise, HE' performs slightly better than the original algorithm.  The original algorithm attempts to optimize on $\Delta Q$ solely but it is known that greedy optimization does not necessarily lead to fully optimized result.  Heuristics HE' is our proof of the fact that CNM algorithm can be improved in both speed and modularity.  It processes \mixiGraph {500K} data set 7 times faster, improves modularity by 8-11\%, and can process much larger data set that are incapable for the original proposal to process.

Heuristics HN performs slightly better in speed than HE but the community structures they produce exhibit rather poor modularity: they were lower than the modularity resulted from the original algorithm by 21-28\%.

\begin {figure}
  \centerline {\includegraphics [width=0.95\linewidth]
    {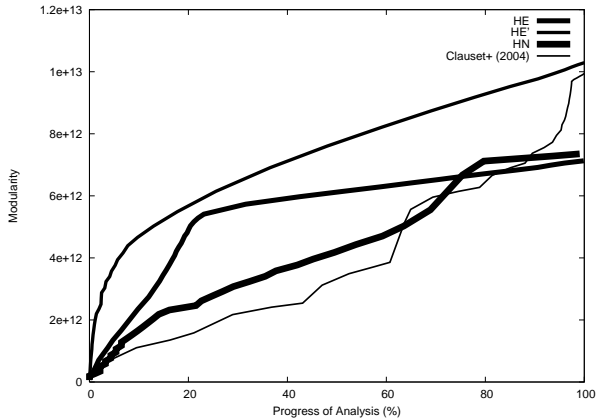}}
  \caption {Growth of modularity as community analysis progresses.  The data set used is \mixiGraph {500K}}
  \figlabel {comparison/modularity-etime}
\end{figure}

It is interesting to see how modularity is improved as the community analysis progresses (\figref {comparison/modularity-etime}).  The horizontal shaft is normalized to the elapsed time of each community analysis.

In the original algorithm, modularity gradually improves.  Though it attempts greedily optimize on modularity, modularities of community structures computed by using heuristics are superior during the first half of the computation.  This chart also suggests that greedy optimization does not successfully optimize modularity.

\begin {figure}
  \centerline {\includegraphics [width=0.95\linewidth]
    {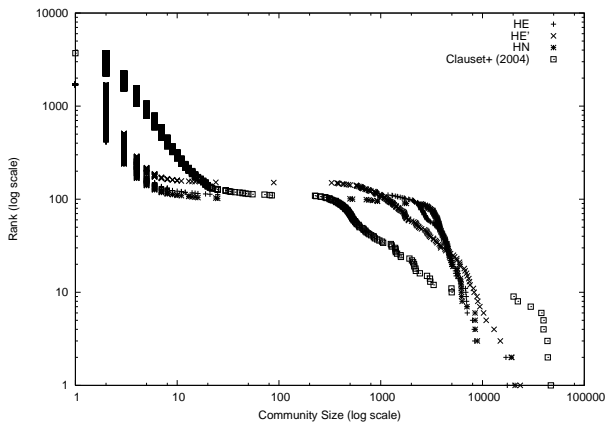}}
  \caption {Sizes of communities: both shafts are in log-scale.}
  \figlabel {comparison/plot500000}
\end {figure}

Heuristics HE' demonstrates steep growth of modularity in the very early stage and it grows steadily up to the end of analysis.  The growth of HN is similar to the original algorithm.  In HE, modularity grows rather steeply but its growth almost stops shortly.  It might be possible to interpret this fact that HE forms core structure in its earlier stage and that we can stop community analysis at the early stage which produces an approximation of the community structure.

So far we have mainly discussed the quality of community clusterings in terms of their modularities as defined in \cite {ng04:modularity}.  It is an important issue to compare the structures produced by four flavors of CNM algorithm.  \figref {comparison/plot500000} depicts a histogram of community size in a log-scale chart.  All methods find a few large ($> 10,000$) communities and a lot of small ($< 10$) ones.  Also they find almost no middle-sized communities.  The original algorithm finds larger communities ($> 20,000$ members) than our heuristics.

An important question to answer is ``existence of correspondence between communities found using different flavor of CNM algorithm''.  If it is not the case, reliability of the results produced by CNM algorithm may need to be reconsidered.  At this moment, this remains to be an open question.

\subsection {Scalability}

\begin {figure}
  \centerline {\includegraphics [width=0.95\linewidth]
    {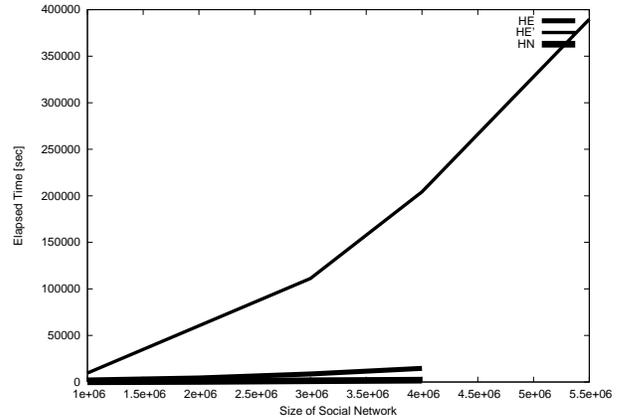}}
  \caption {Scalability}
  \figlabel {comparison/etime-size-M}
\end {figure}

\figref {comparison/etime-size-M} is obtained from applying proposed heuristics on larger data sets, ranging from 1M nodes up to 5.5M nodes.  HE and HN demonstrates almost linear speed up.  Scalability of HE', on the other hand, is slowly declining but we estimate that it is applicable to networks that has up to 10M nodes.  Scalability of the algorithm is bound by memory size for standard PC.  HE and HN failed to process a network that consists of 5.5M nodes due to lack of physical memory.

Our current implementations of CNM algorithm are not optimized for reduction of memory usage.  We plan to re-implement it and achieve better use of memory.  Hopefully we achieve to analyse larger networks with 10M nodes, soon.  Further acceleration of the algorithm requires use of parallelism.


\section {Summary}
\seclabel {summary}

The paper identified a bottleneck of a community analysis algorithm proposed by Clauset, Newman, and Moore \cite {cnm04:very_large_network}.  Its inefficiency was caused from unbalanced structuring of communities.  The paper proposes three heuristics that attempt to balance the size of communities being merged.  We have removed the bottleneck and successfully obtained community structures of large scale social networks that contain over 5,000,000 nodes.  Our approach is scalable.  It is expected to scale to a SNS network that contains 10,000,000 nodes.

There still remain unanswered interesting issues.  How are community structures found by different algorithm relate with each other?  How algorithmically found cyber-community structures relate to human communities.  Is it possible to explain the dynamics of SNS community growth in terms of the progress of community analysis?

From a technical stand point, we are interested in how much faster and how scalable are our proposals.  We are interested in parallelization of community analysis.  The impact of our research to middle-scale  social network is large.  Our research has made it possible to analyse a middle scale social network (with 100,000 nodes) in a few minutes on a standard laptop computer and we are freed from waiting response from community analysis performed on a server for days and hours.

We are currently working on visual presentation of cluster structures with Dr.~Hiroshi Hosobe and Mr.~Minato Koshida.  We are also working on analysis of cyber-communities found in social networking services and their dynamics.

\bibliographystyle {hplain}
\bibliography {papers}

\end {document}